%% file: main.tex
\pdfoutput=1

\documentclass[nonacm,sigplan]{acmart}
\usepackage{parskip}
\usepackage{xspace}
\usepackage{tabularx}
\usepackage{multirow}
\usepackage{kotex}
\usepackage[ruled,vlined,linesnumbered]{algorithm2e} 
\usepackage[]{hyperref}
\usepackage[title,titletoc]{appendix}
\usepackage{etoolbox}
\BeforeBeginEnvironment{appendices}{\clearpage}
\usepackage[]{hyperref}

\usepackage{algpseudocode}

\newcommand{\sysname}{ELIS\xspace}
\newcommand{\abb}[1]{{\textsf{\small{#1}}}\xspace}

\newcommand{\compname}{Samsung\xspace SDS\xspace}
\newcommand{\svcname}{FabriX\xspace}
\newcommand{\schedname}{\textsc{ISRTF}\xspace}

\begin{document}

\title{\sysname: Efficient LLM Iterative Scheduling System with Response Length Predictor}

\author{Seungbeom Choi*, Jeonghoe Goo*, Eunjoo Jeon, Mingyu Yang, Minsung Jang }
\affiliation{%
  \institution{Cloud Research Team, Samsung SDS}
   \country{South Korea}
  }

\thanks{* indicates equal contribution}
\thanks{ Correspondence to: Minsung Jang <minsung.jang@samsung.com>.}

\input{0_abstract}

\maketitle 
\pagestyle{plain} 

\input{1_introduction}

\input{2_background}
\input{3_motivation}
\input{4_design}
\input{5_implementation}
\input{6_evaluation}
\input{7_related_work}
\input{8_conclusion}

\input{9_ack}
\bibliographystyle{plain}

\bibliography{10_references}
\input{N_Appendix}
\end{document}

%% file: 0_abstract.tex
\begin{abstract}

We propose \sysname, a serving system for Large Language Models (LLMs) featuring an \textit{Iterative Shortest Remaining Time First (ISRTF)} scheduler designed to efficiently manage inference tasks with the shortest remaining tokens.
Current LLM serving systems often employ a first-come-first-served scheduling strategy, which can lead to the “head-of-line blocking” problem.
To overcome this limitation, it is necessary to predict LLM inference times and apply a shortest job first scheduling strategy.
However, due to the auto-regressive nature of LLMs, predicting the inference latency is challenging.
\sysname addresses this challenge by training a response length predictor for LLMs using the BGE model, an encoder-based state-of-the-art model.
Additionally, we have devised the \textit{ISRTF} scheduling strategy, an optimization of shortest remaining time first tailored to existing LLM iteration batching.
To evaluate our work in an industrial setting, we simulate streams of requests based on our study of real-world user LLM serving trace records.
Furthermore, we implemented \sysname as a cloud-native scheduler system on Kubernetes to evaluate its performance in production environments.
Our experimental results demonstrate that \textit{ISRTF} reduces the average job completion time by up to 19.6\%.

\end{abstract}

%% file: 1_introduction.tex

\section{Introduction}

Since the launch of OpenAI ChatGPT in 2022, numerous Large Language Models (LLMs) and services have been introduced, achieving high--performance~\cite{nips:gpt3, arxiv:llama3}. LLMs based on Transformer decoders differ from traditional machine learning and encoder--based models in several ways. First, LLMs can perform multiple tasks such as summarization, text generation, and question--answering with a single model, meaning that one LLM must handle various user requests. Second, LLMs generate tokens auto--regressively, often causing longer inference latency compared to Transformer encoder-based models like BERT, which can process tokens in parallel. Third, LLMs are GPU memory-bound during inference, leading to speed degradation due to their billions of parameters.

Despite the challenges mentioned above, it is essential to serve as many user requests as possible at minimal cost.
To achieve this goal, user requests for LLM services arriving at different times are batched together and passed to the LLMs. This traditional batching process creates inefficiency, as short-context jobs must wait for longer ones to finish. ORCA addresses this problem by implementing iteration-level batch scheduling, where jobs are processed per iteration without waiting for the entire batch to complete~\cite{osdi:orca}.

\begin{table}
  \caption{Comparison to prior works.}
  \label{table:accuracy}
  \begin{tabular}{lccc}
  \toprule
  \textbf{Related Work} & \textbf{Scheduling} & \textbf{Predict} & \textbf{Preemption} \\
  \midrule
   ORCA \cite{osdi:orca} & FCFS & X & X \\
   Zheng \textit{et al.}~\cite{nips:PiA} & SJF & One-off & X \\
   \textit{ELIS} & \textit{SRTF} & \textit{Iterative} & \textit{O} \\
  \midrule
  \end{tabular}
\vspace{-0.2in}
\end{table}

However, iteration-level batching uses a First--Come--First--Served (FCFS) scheduling strategy, which is non-preemptive and can lead to head-of-line blocking. Designing optimal latency-based priority scheduling for LLMs is challenging for several reasons. First, decoder-based LLMs have difficulty predicting the length of the generated output tokens, which is a major factor in latency. Second, the latency of the predictor should not become a bottleneck; thus, models that require a relatively large amount of computing resources, such as decoder-based generative models, cannot be used despite their high accuracy. Third, applying latency-based priority scheduling involves changing the priority of ongoing jobs and managing high context change costs, including Key-Value (KV) cache changes, which complicates the process.

Research on serving schedulers for LLMs has focused on enhancing iterative batching by predicting the execution time of each step and performing jobs with shorter execution times first using the Shortest Remaining Time First (SRTF) scheduler. Zheng \textit{et al.}\ conducted instruct fine--tuning so that LLM models could provide both their responses and the length of their responses simultaneously~\cite{nips:PiA}. By executing jobs with shorter response lengths first, they achieved over an 80\% increase in throughput. However, a critical drawback of this approach is that instruct fine-tuning the LLM model can affect the original model’s accuracy. Qiu \textit{et al.}\ used a BERT model to predict the inference speed of LLMs and processed requests with a Shortest Job First (SJF) scheduler~\cite{aiops2024qiu}. Their results showed a 39.1\% reduction in total Job Completion Time (JCT) compared to FCFS and a 2.21-fold increase in throughput. However, the BERT predictor introduced in the study had low accuracy with an F1-score of 0.6 and did not have a fallback plan for incorrect LLM inference time predictions, which could still potentially lead to head-of-line blocking.

To address these challenges, we propose \sysname, an \textbf{E}fficient \textbf{L}LM \textbf{I}terative \textbf{S}cheduling system designed to reduce average JCT by efficiently scheduling tasks based on predicted response lengths.
The distinct features of our research are as follows:

First, we developed an SRTF-based LLM request scheduler, \textit{Iterative Shortest Remaining Time First} (\textit{ISRTF}), using a response length predictor. ISRTF prioritizes tasks with shorter remaining times, where the remaining time is predicted over several iterations using the method described in Section~\ref{sec:predictor}. We designed a modular architecture for the predictor, allowing the scheduler to operate in a model-agnostic manner. As the prediction performance improves through retraining based on log data, the effectiveness of SRTF can also increase. In this study, an iterative predictor was developed and applied ($R^2 = 0.852$), and we confirmed that accuracy increases with each iteration (Section~\ref{subsec:iterative_predict}). When applying the predictor, the ISRTF scheduler reduced the average JCT by up to 19.58\% compared to FCFS on NVIDIA A100 GPUs.

Second, we implemented \sysname at the production level based on industrial workloads.
In this study, we extracted trace distributions based on two months of actual operation data from \svcname\footnote{Full name to be disclosed in final version of the paper.}, a cloud-native LLM service managed by our organization. Using these traces, we developed a request generator based on a Gamma distribution. Additionally, \sysname was implemented with Kubernetes, the most widely used container orchestrator for production, leveraging its pod auto-scalability and reliability features. We conducted performance studies using vLLM~\cite{SOSP:vLLM}, a commonly utilized execution engine in both academic and industry research. \sysname successfully distributed prompts across more than 10 nodes and displayed near-linear scaling performance, achieving 18.77 Requests Per Second (RPS) with 50 workers on NVIDIA H100 GPUs.

Third, from the standpoint of a cloud service provider, GPUs are finite resources. 
Therefore, stopping resources allocated to lower-priority jobs and prioritizing high-priority LLM requests can enhance GPU utilization efficiency.
In this study, we investigate the efficacy of preempting tasks using trace data obtained while our organization was hosting \svcname, and we report our results in Section~\ref{subsec:preemption}.
We have also included the code for adjusting the frequency of preemption in the public code of \sysname.

The primary contribution of this paper is the introduction of \sysname, an iterative priority-based LLM task serving system, which reduced the average JCT by up to 19.58\% compared to FCFS by deploying ISRTF scheduling. In addition, we contribute to the research community by implementing \sysname with the vLLM execution engine as Kubernetes components, enabling easy deployment.

%% file: 2_background.tex
\section{Background}
In this section, we explain the architecture of LLMs, focusing on the auto-regressive inference process, which is divided into two distinct phases: prefill and decode. We also describe strategies for efficient LLM serving, such as batching and scheduling.

\subsection{LLM Inference Process}
\paragraph{Autoregressive Generation:} 
Typically, the inference procedure of LLMs can be divided into two distinct phases: prefill and decode. In the prefill phase, which corresponds to the first iteration, the prompt is processed to generate the KV cache. In the decoding phase, the query, key, and value of new tokens are calculated step by step and added to the KV cache generated during the prefill phase~\cite{osdi:distserve}. 
    
The latency of LLM inference can be characterized by two metrics. One is the \textit{Time To First Token (TTFT)}, which is the duration of the prefill phase, and the other is the \textit{Time Per Output Token (TPOT)}, which represents the average time taken to generate each token during the decoding phase~\cite{osdi:distserve}. Total LLM latency can be calculated as \textit{TTFT} + (\textit{TPOT} $\times$ number of tokens to be generated). Although the LLM latency formula is clear, it is difficult to predict the latency of each request. This is because, due to the auto-regressive nature of LLMs, it is hard to predict the number of output tokens that will be generated for a given prompt, which has the greatest impact on LLM latency.

\subsection{Efficient LLM Serving Strategies}
LLMs such as GPT-3~\cite{nips:gpt3} and PaLM have been deployed in services like Bing and Bard, handling a large number of requests every day. 
Providing satisfying service to users while reducing the inference cost of LLMs becomes a crucial issue~\cite{nips:PiA}. 
Over the years, significant developments have been made in efficient LLM serving techniques, such as kernel fusion, pipeline parallelism, tensor parallelism, and quantization.
In this section, we explain LLM inference batching and request scheduling methods to reduce LLM inference costs.

\paragraph{Continuous Batching:} LLM services perform inference by batching inputs from multiple users at different intervals to increase throughput. 
Conventional batching starts a batch all at once and waits for all requests in the batch to complete their computation.
In contrast, ORCA introduces \textit{iteration-level batching}, also known as \textit{continuous batching}, which does not wait for all requests to complete before starting a new batch.
Instead, it continuously schedules a new request when a request in the processing batch completes and GPU slots are available~\cite{osdi:orca}. 
However, because it uses a non-preemptive scheduling method, head-of-line blocking issues may occur, making it unsuitable for interactive services using LLMs~\cite{aiops2024qiu}.

\paragraph{Priority Scheduling:} SJF scheduling has the advantage of guaranteeing minimum average waiting time compared to FCFS, by scheduling tasks with shorter execution times first.
The execution time of convolutional neural networks or encoder-based models like BERT can be easily predicted based on the model size~\cite{osdi:clockwork, acm:inferline}.
However, this observation does not hold true for LLMs due to the non-deterministic nature of the auto-regressive decoding phase included in LLMs.
FastServe attempted to reduce the average JCT by utilizing a Multi-Level Feedback Queue (MLFQ)-based scheduling approach~\cite{arxiv:fastserve}.
MLFQ is less efficient because it needs to adjust the priority of each task through trial and error, causing head-of-line blocking in the process.
Qiu \textit{et al.}~\cite{aiops2024qiu}, who presented the most similar work, used BERT to predict the output token length of LLMs, which was used for priority and scheduling requests.
However, predictions were made only once, making it difficult to fully capture the auto-regressive characteristics of LLMs.

%% file: 3_motivation.tex
\section{Motivation}

In this section, we describe the main intuitions of our work and introduce the rationales behind our design choices. 

\subsection{Ability to Comprehend Natural Language}

To predict both the content and length of the generated answers according to the context, it is crucial that the embedding model understand the context of prompts.
Hence, we pose the question: \textit{``Can the embedding model represent the context of the user prompts?''} and aim to provide a valid answer in this section. 

In this paper, we deploy the BGE (BAAI/bge-base-en-v1.5) model~\cite{bge_embedding} as the base architecture for our LLM response length predictor. The BGE model is the state-of-the-art on the Massive Text Embedding Benchmark leaderboard (ranked first in July 2024; ranked second in October 2024).

We validated the effectiveness of the BGE model through the following experiment. Using ChatGPT, we generated two datasets: one consisting of 100 semantically similar sentences and another consisting of 100 semantically unrelated sentences. The similar sentences focused on the topic of weather, while the unrelated sentences covered various topics not related to weather. To ensure there was no overlap in vocabulary or duplicate sentences between the two datasets, we applied thorough pre-processing steps, including removing any sentences with shared words or identical content. The CLS token vector of each sentence was then extracted using the BGE model, resulting in 768-dimensional embeddings. These embeddings were reduced to two dimensions using principal component analysis for visualization.

\begin{figure}
    \centering
    \includegraphics[width=1.0\linewidth]{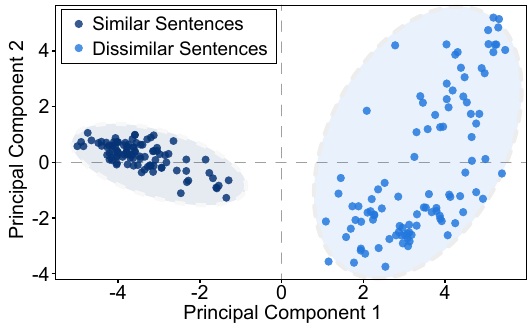}
    \caption{BGE CLS vector distance with different groups.}
    \label{fig:CLS vector distance}
\end{figure}

Figure~\ref{fig:CLS vector distance} illustrates that sentences with similar contexts (colored dark blue) are clustered closely together, whereas sentences with dissimilar contexts (colored light blue) are scattered farther apart. This observation confirms that the BGE model effectively captures the contextual information of sentences in the CLS token embeddings.

\subsection{Capability of Predicting Response Length}

We conducted the following experiment to assess the capability of BGE to predict response length:

The architecture of our response length prediction model consists of the pre-trained BGE model~\cite{bge_embedding} and eight additional linear layers. In our research, we focused on the CLS token, which is known to contain significant input information~\cite{arxiv:bert}. The input prompt is processed by the BGE model to generate a CLS token embedding, which is then fed into the subsequent linear layers to predict the expected response length. The pre-trained parameters of the BGE model were frozen, and only the eight linear layers were trained during fine-tuning for output token length prediction.

We used the LMSYS-Chat-1M dataset for training~\cite{zheng2023lmsyschat1m}, which was collected from the Vicuna demo and the Chatbot Arena website. This dataset contains one million chat records from 25 state-of-the-art LLMs, with 49\% of the data generated by the Vicuna-13B model and the remainder from 24 other models. Since our focus in this section is to evaluate the feasibility of using BGE as a predictor, we did not apply any bias mitigation techniques.

\begin{table}
  \caption{BGE baseline prediction results.}
  \label{table:BGE baseline prediction results}
  \centering
  \begin{tabular}{lccc}
  \toprule
  \textbf{Model}& \textbf{MAE}& \textbf{RMSE} & \textbf{$R^2$} \\
  \midrule
   Pre-trained BGE & 175.99 & 224.98 & -1.58 \\
   Fine-tuned BGE   & 71.48  & 101.29 &  0.48 \\
  \midrule
  \end{tabular}
\end{table}

Table~\ref{table:BGE baseline prediction results} shows the results after fine-tuning the output length predictor with the LMSYS dataset. The evaluation metrics used were Mean Absolute Error (MAE), Root Mean Square Error (RMSE), and R-squared ($R^2$).
The fine-tuned BGE model demonstrated lower MAE and RMSE values compared to the pre-trained BGE, and its $R^2$ value was closer to one.
Performance improvement was observed despite data imbalance and the use of default training parameters.
This suggests that even higher performance could be achieved if the model were trained with data specifically designed for LLM scheduling.
Qiu \textit{et al.}\ found that when the accuracy of the output token predictor was 0.615, JCT decreased by 39\% and throughput increased by 2.2 times~\cite{aiops2024qiu}.

Therefore, given that our fine-tuned BGE model achieves a high prediction accuracy, it is expected to provide significant performance benefits as an SRTF scheduler.

\subsection{Improving Reliability with Iterative Prediction}
\label{subsec:iterative_predict}

The key intuition of this study is that, for LLMs, output is generated in \textit{iterations} due to their auto-regressive nature. ORCA has designed iteration-level batching methods, such as continuous batching and in-flight batching, to accommodate this process.
The question we pose in this section is: \textit{``If the LLM delay predictor receives input in iterations, will the accuracy improve as more information is provided to the predictor incrementally?''}

\begin{figure}
    \centering
    \includegraphics[width=1\linewidth]{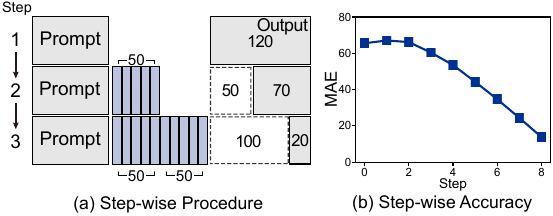}
    \caption{(a) Illustration of prediction procedure where each step (iteration) comprises of 50 tokens and (b) MAE of predictor for each step.}
    \label{fig:iteration_process}
\vspace{-0.1in}
\end{figure}

Figure~\ref{fig:iteration_process}(a) depicts a graphical illustration of performing prediction in iterations.
Let us assume that the generated output is 120 tokens in length when given a user prompt.
In Step 1, the predictor should estimate an output length of 120 tokens, receiving only the user prompt as input.
In Step 2, the predictor adjusts its estimate to 70 tokens, using both the prompt and the previously generated 50 tokens as input.
As the process continues, the previously generated tokens are continuously fed back into the input, and the corresponding number of tokens is excluded from the output prediction, refining the estimate at each iteration.
We have empirically determined that the optimal window size is 50 tokens through several experiments.

As shown in Figure~\ref{fig:iteration_process}(b), the MAE of the LLM latency predictor decreases as the iterations progress.
Therefore, we conclude that the accuracy of our predictor can be improved when provided with partial output at every iteration.

\subsection{Preemption and Real-world LLM Serving}
\label{subsec:preemption}

Preemption is a widely studied scheduling method for ensuring priority among multiple tasks running on a system with limited resources.
By preempting a lower-priority task and reallocating resources to higher-priority tasks, the higher-priority tasks are more likely to finish sooner. 
As a result, preemption has been widely researched and deployed in various machine learning systems~\cite{nsdi:shepherd,eurosys:lyra,eurosys:orion,eurosys:varuna,SOSP:vLLM}, including vLLM, which is the inference engine used in \sysname.

According to our experimentation on vLLM with real-world prompts sampled from the LMSYS-Chat-1M dataset, we discovered that the probability of preempting a task due to limited resources is relatively low.

We conducted experiments on our prototype system to observe the minimum batch size that can induce preemption due to limited memory for storing the KV cache.
These experiments were performed on an NVIDIA A100 GPU.
Details and results of our experimentation are reported in Appendix~\ref{sec:append_data}.
LLaMA2-13B, being a relatively large model, has a higher likelihood of preemption and began preempting requests at a batch size of 120.
Combining this with the average latency reported in Table~\ref{tab:ml-models}, the average request rate required to constantly saturate this batch size is 13.9 requests per second ($=\frac{120}{8.61}$).
Given that the maximum daily request rate observed from \svcname---an LLM-based service used by employees of \compname\footnote{Organization will be disclosed in the final version of the paper.}---is below 3 requests per second, we can conclude that the probability of requiring preemption when serving LLM inference tasks is relatively low.
Additionally, according to a survey on public Azure Machine Learning Service~\cite{arxiv:burstgpt} trace data, the maximum rate did not exceed 2 requests per second.
Hence, we did not include further experiments for preemption and instead focused on iterative priority scheduling.
Nonetheless, we have designed and implemented policies that can adjust the frequency of preemption and prevent starvation for future research.
All implementations will be included in the public code of \sysname.

%% file: 4_design.tex
\section{Design}
\label{sec:design}
\begin{figure}
    \centering
    \includegraphics[width=1.0\linewidth]{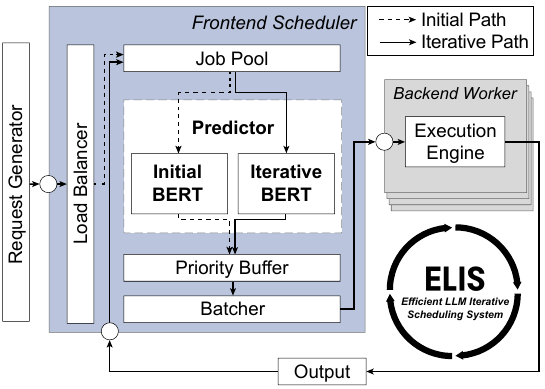}
    \caption{Overall architecture of \sysname.}
    \label{fig:overall_architecture}
\vspace{0.2in}
  \end{figure}

Figure~\ref{fig:overall_architecture} illustrates the overall architecture of \sysname, which consists of a request generator, a frontend scheduler that includes a predictor, and multiple backend workers. The following subsections provide a description of how each component interacts with the others and details regarding our predictor model.

\setlength{\algomargin}{2em} 
\begin{algorithm}[]
	\small
	\LinesNumbered
	\SetAlgoLined
 
	\KwIn{Set of Prompts $P$, Global State $G$}
	\KwOut{Response of each prompt in $P$}
        
        \For{$prompt$ \emph{in} $P$}
        {
            store the text of $prompt$ in a new $job$; 
 
	    $job$.$node$ $\leftarrow$ $Load Balancer$.get\_min\_load($G$); 

            $Job Pool$.push($job$);
        }
       
	\For{\emph{every} $iteration$}{

            \If{$Job Pool$ \emph{is} \emph{empty}}{
                end $iteration$ loop;
            }

            \For{$job$ \emph{in} $Job Pool$}{

                \eIf{$job$.$priority$ \emph{is} \emph{none}}
                {$job$.$priority$ $\leftarrow$ $Predictor$.init($job$);}
                {$job$.$priority$ $\leftarrow$ $Predictor$.iter($job$);}
                
                pop $job$ from $Job Pool$;

                push $job$ to $Priority Buffer$;
            }
            
            $Batched Prompt$ $\leftarrow$ $Batcher$.batch($Priority Buffer$);
            
            $output$ $\leftarrow$ $Backend Worker$.exec($Batched Prompt$);

            \For{$job$ \emph{in} $output$}{

                \eIf{$job$.$response$ \emph{is} \emph{finished}}
		      {return $job$.$response$;}
                { 
                add $response$ to $job$.$response$;
                
                push $job$ back to $Job Pool$; }
            
            }

	} 
	\caption{Overall scheduling flow of \sysname.}
	\label{alg:scheduler_flow}
\end{algorithm}

\subsection{Overall Scheduling Flow of \sysname}
\label{sec:scheduler}

\sysname deploys a central frontend scheduler responsible for prioritizing prompts and assigning appropriate backend workers for execution.
The scheduler makes scheduling decisions when new prompts arrive or when a scheduling iteration has completed and a new iteration begins.
Each \textit{scheduling iteration} processes \textit{50} tokens of a batched prompt.
Therefore, whenever a new decision is made, it influences the next iteration, which processes the next set of \textit{50} tokens.

\paragraph{Scheduling Process:}
Algorithm~\ref{alg:scheduler_flow} describes the overall scheduling procedure of \sysname, which receives a set of prompts $P$ and returns the response for each prompt.
Upon the arrival of a prompt, the frontend scheduler converts the prompt into a $job$, which is a data record managed internally by the scheduler \textit{(line 2)}.
The job is then assigned to a backend worker node by the load balancer.
The load balancer greedily distributes the jobs among the worker processes.
By consulting the global state $G$ stored in the frontend, which includes the number of jobs running on each backend worker, the load balancer selects the worker executing the fewest number of jobs (\textit{line 3}).
After the backend worker is selected, the new job is pushed to the $Job Pool$, which is a queue that stores $job$s (\textit{line 4}).

In each iteration, scheduling decisions are made until there are no $job$s remaining in the $Job Pool$ (\textit{lines 6 to 9}).
Based on the scheduling policy and whether a $job$'s priority has been previously assigned, each $job$ is given a priority and pushed to the $Priority Buffer$ (\textit{lines 10 to 18}).
The $Priority Buffer$ consists of multiple priority queues, where each queue stores $job$s assigned to a specific $node$.
Whenever a backend server becomes available, a batched prompt is formed, starting with the prompt with the highest priority (\textit{line 19}).
After forming a batched prompt, it is sent to the backend engine and executed for one window of \textit{50} tokens (\textit{line 20}).
The frontend scheduler receives the $output$ after one window, which includes a list of jobs and their corresponding $responses$. 
Each $job$ in the $output$ is checked to determine whether the response is complete (\textit{lines 21 to 22}).
If the $response$ is complete, the full response is returned (\textit{line 23}).
If not, the partial response is stored in the $job$, and the $job$ is pushed back to the $Job Pool$ to be processed in the next scheduling iteration (\textit{lines 24 to 26}).

\paragraph{Performance Optimization:}

We have implemented additional optimizations to \sysname to enhance performance and scalability.
The scheduling process is divided into several sub-procedures, each executed asynchronously.
Each procedure maintains a buffer for sharing and communicating data with the others.

To avoid overburdening the network, the input prompt of each job is sent to the backend only once, rather than being sent for every scheduling iteration.
Batching is performed using fixed window sizes of \textit{50} tokens, and partial outputs are sent in these batches.

\paragraph{Real-World Request Analysis:}
\label{subsec:workload}

\begin{figure}
    \centering
    \includegraphics[width=1.0\linewidth]{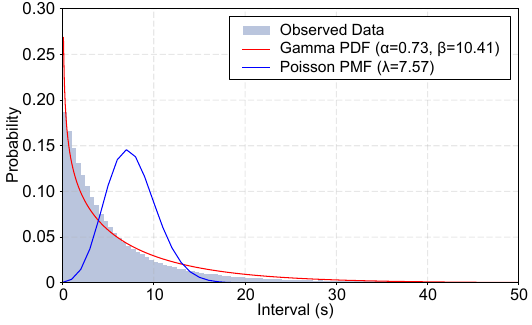}
    \caption{Request interval distribution of LLM serving. The Gamma PDF and Poisson PMF distributions were fitted based on the observed data.}
    \label{fig:Fabrix_with_distribution}
\vspace{-0.2in}
\end{figure}

Due to the lack of publicly available trace data for LLM services, most existing studies on LLM serving have assumed that request arrival times follow a Poisson process or have utilized Azure function trace data~\cite{nips:s3, aiops2024qiu, arxiv:fastserve, osp:pagedattention}. 
However, in this study, we analyzed over 200,000 real-world trace data points gathered over two months from \svcname, an LLM service operated by \compname.
We confirmed that the intervals between LLM requests follow a Gamma distribution more closely than a Poisson distribution, with shape parameter $\alpha = 0.73$ and scale parameter $\beta = 10.41$, as shown in Fig.~\ref{fig:Fabrix_with_distribution}.
This result is similar to BurstGPT~\cite{arxiv:burstgpt}, which chose the Gamma distribution to better capture the burstiness of LLM service requests.
In conclusion, we randomly sample requests from a Gamma distribution when evaluating \sysname in Section~\ref{sec:evaluation}.

\paragraph{Backend Worker:} 

Each backend worker of \sysname is responsible for relaying inputs to the inference engine, ensuring that the designated priority is maintained when executing input prompts and sending replies generated by the execution engine.
In other words, the backend worker acts as a proxy to the inference engine.
We chose vLLM~\cite{SOSP:vLLM} as our execution engine due to its popularity in both industry and academia.

To successfully control batched prompts in iterations, we have added two additional features to vLLM.
The first feature is \textit{iteration-wise execution}, which executes a batched prompt for \textit{K} tokens (the window size of an iteration).
A batched prompt passed to the vLLM engine is executed and returns partial responses when all prompts in the batch have produced \textit{K} tokens or when a prompt has finished.
The partial responses are sent to the frontend scheduler, and the backend worker checks if any pending commands have been sent by the frontend scheduler.
If there are pending batches, the backend server checks and updates the next batch to be executed by vLLM.
The second feature is \textit{configurable priorities}, which is used to override the default priority of vLLM (FCFS).
Whenever the backend server updates the next batch, the priority of each prompt is also updated.
The priority assigned to each prompt affects the order in which tasks are preempted by vLLM.
Whenever preemption is required, vLLM will check the priority of each task and preempt the tasks starting with the lowest priority.
Please refer to Section~\ref{sec:implementation} for additional details on how we have implemented each feature.

\subsection{Response Length Prediction Model} 
\label{sec:predictor}
  
In this study, we tested an online scenario by running vLLM as a backend worker with the ISRTF scheduler on Kubernetes. Output from 13 LLMs (listed in Table~\ref{tab:13model_list}), including LLaMA1, LLaMA2, Vicuna, OPT, and GPT-NeoX, was collected by executing them on the vLLM framework. 11,000 prompts were randomly selected from the LMSYS-Chat-1M dataset, and the LLMs were run with the vLLM default settings to generate a total of 143,000 prompt-answer pairs. The collected data consist of model name, input tokens, output tokens, input token length, output token length, and execution time.
During the dataset construction process, duplicate data were initially removed, and outliers were then removed using the Interquartile Range and log transformation methods. In this process, approximately 40,000 rows were removed, leaving 105,295 rows.

We treated the user's original prompt and the input consisting of the prompt attached with the answer as separate data. Therefore, step data for iteration prediction was prepared for each prompt. The entire dataset was shuffled and then divided into training, validation, and test sets in a 6:2:2 ratio, respectively.

The model architecture consists of the BGE model and eight Fully Connected (FC) layers. The CLS token and other token embeddings generated by the BGE model are passed through mean pooling, and the expected response length is predicted using eight FC layers. The activation function used was ReLU, and the hidden dimension of the FC layers was set to 1024. The learning rate used was $1 \times 10^{-4}$. Training was performed on an NVIDIA A100 GPU with a batch size of 16, and the loss converged after epoch 16. The results of the BGE fine-tuning on the vLLM dataset showed that the MAE was 19.923, the RMSE was 34.327, and the $R^2$ was 0.852. Additionally, the $R^2$ value increased by a factor of 1.78 compared to the fine-tuned BGE on the LMSYS dataset.

%% file: 5_implementation.tex
\section{Implementation}
\label{sec:implementation}

\paragraph{System Software Architecture:}
As illustrated in Figure~\ref{fig:overall_architecture}, our multi-GPU LLM serving system comprises a frontend scheduler server and multiple backend workers, each deploying a vLLM execution engine~\cite{SOSP:vLLM}.
The serving system is developed in Python with approximately 3.1K lines of code, including both the frontend scheduler and backend worker implementations.
The frontend scheduler consists of multiple software modules, each instantiated as a separate process.
Each process collaborates through interprocess communication using shared memory, which stores the global state of each backend engine and the jobs running in the serving system.
Each backend worker is responsible for executing inference for a specific model with vLLM, determined by loading a designated file (a pickled zip file with the extension \texttt{.pth}) when initiating the worker.
We have made minor adjustments to the policy, scheduler, and entry points of vLLM (less than 100 lines of code) to execute batched prompts \textit{K} tokens at a time and override the default preemption policy, which is FCFS.
We have added a custom policy class to implement SRTF scheduling. This custom policy allows us to prioritize jobs based on their remaining execution time.
The entry point was modified to return the request ID managed internally by vLLM.
It is necessary for the worker to manage the mapping between jobs and request IDs, as the priority of each job will be updated using the internal request ID.

\paragraph{Deploying \sysname on Kubernetes:}
We have deployed our prototype as components within a Kubernetes cluster.
The frontend scheduler and backend execution servers are deployed as pods within the Kubernetes cluster.
Given the widespread use of Kubernetes for managing microservices, pods are treated as ephemeral resources that can be automatically replaced when necessary.
However, each backend pod must be uniquely identifiable, as the frontend pod needs to communicate with the specific pod assigned to execute a batch.
To achieve this, the backend pods are managed using a StatefulSet, ensuring each pod has a unique ID for proper identification and communication.
Communication between the pods is facilitated through endpoints exposed as services within the Kubernetes environment.
The YAML configuration files defining the Kubernetes components used in the experiments presented in this paper will be made publicly available.

%% file: 6_evaluation.tex
\section{Evaluation}
\label{sec:evaluation}

\subsection{Methodology}

\newcolumntype{L}[1]{>{\raggedright\arraybackslash}p{#1}}
\newcolumntype{C}[1]{>{\centering\arraybackslash}p{#1}}
\newcolumntype{?}{!{\vrule width .07em}}

\begin{figure*}
    \centering
    \includegraphics[width=1.0\linewidth]{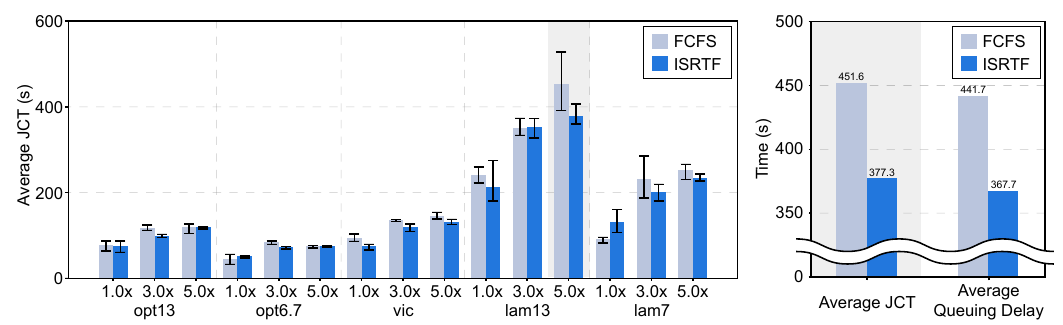}
    \caption{\textsc{(left)} \textit{JCT} comparison between \textsc{FCFS} and \schedname where each experiment uses a multiple of average throughput. Bar represents the average value and each tick represents the minimum and the maximum value of each experiment. \textsc{(right)} Average \textit{JCT} and queuing delay of \abb{lam13} with 5.0x RPS (case highlighted in gray shading).}
    \label{fig:aver_jct_per_model}

\end{figure*}

\paragraph{Experimental Setup:}
We have evaluated \sysname by deploying the prototype system on a Kubernetes cluster hosted on a server from our organization's private infrastructure.
Table~\ref{tab:system-specifications} provides the specifications of the server's software components and GPU.

To avoid network bottlenecks when generating requests from outside the cluster, the prompts are generated by the frontend scheduler.
Nevertheless, we have included a stand-alone generator in our public code for future research.

\paragraph{Baseline Scheduling Algorithms:}

\begin{table}[t]
\caption{The evaluated system specifications.}
	\centering
		\resizebox{0.48\textwidth}{!}{
			\footnotesize
			\begin{tabular}{L{10em}L{15em}}
			    \hline
			    \multicolumn{2}{c}{\textbf{System Overview}} \\
				\hline
				\textbf{CPU} & Dual AMD EPYC 7H12 64-core \\
				\textbf{GPU} & 8 NVIDIA A100 \\
				\textbf{Memory Capacity} & 2 TB \\
				\textbf{Operating System} & Ubuntu 20.04  \\
				\textbf{CUDA} &  12.2 \\
				\textbf{NVIDIA Driver} & 535.183 \\
				\textbf{ML framework} & vLLM v0.5.0 \\
				\hline
			    \multicolumn{2}{c}{\textbf{GPU Specification}} \\
				\hline
				\textbf{CUDA Cores} & 6,912  \\
				\textbf{Memory Capacity} & 80 GB HBM2 \\
				\textbf{Memory Bandwidth} &  2,039 GB/sec \\
				\hline
			\end{tabular}
		}
        
		\label{tab:system-specifications}
\end{table}
\begin{table}[t]
\caption{List of LLM models (and abbreviation) used in the evaluation.}
	\small
	\centering
	\resizebox{0.48\textwidth}{!}{
	\begin{tabular}{L{11em}C{4em}C{4em}}
		\hline
		\multicolumn{1}{c}{\textbf{Model}} & \multicolumn{1}{c}{\textbf{Parameter Size }} & \multicolumn{1}{c}{\textbf{AVG. Latency (ms)}} \\
		\hline
            OPT-6.7B (\abb{opt6.7})~\cite{zhang2022opt} & 6.7B & 1315.5 \\
            OPT-13B (\abb{opt13})~\cite{zhang2022opt} & 13B & 2643.2 \\
            LlaMA2-7B (\abb{lam7})~\cite{touvron2023LlaMA2openfoundation} & 7B & 6522.2 \\
            LlaMA2-13B (\abb{lam13})~\cite{touvron2023LlaMA2openfoundation} & 13B & 8610.2  \\ 
		Vicuna-13B (\abb{vic})~\cite{zheng2023judgingllmasajudgemtbenchchatbot} & 13B & 2964.9 \\
		\hline
	\end{tabular}
	}
	
	\label{tab:ml-models}
	\vspace{-0.1in}
\end{table}

We have ported two priority schedulers into \sysname to compare performance. The two task scheduling algorithms are \textsc{FCFS} and \textsc{SJF}, with \textsc{SJF} serving as an oracle scheduler to indicate ideal performance. \textsc{FCFS} assigns priority based on the prompt's arrival time, while \textsc{SJF} assigns higher priority to prompts with shorter profiled latency. We compare the two schedulers above with our proposed scheduling method, ISRTF.
\paragraph{Large Language Models:}
We have used the models listed in Table~\ref{tab:ml-models} for evaluation. We have prepared \textbf{five} LLMs and report the average latency of 500 prompts executed on an NVIDIA A100 GPU.
\paragraph{Simulated Workload:}
To evaluate \sysname under various workloads, we simulate streams of prompts.
The prompts are sampled from the LMSYS dataset~\cite{zheng2023lmsyschat1m}.
The interval between prompts is randomly sampled from a Gamma distribution in accordance with the conclusion of Section~\ref{subsec:workload}.

\subsection{LLM Serving Performance Comparison}
\label{sec:eval_perf_comparison}
This section compares how well \textit{\schedname} performs compared to other schedulers, \textsc{FCFS} and \textsc{SJF}.

\paragraph{JCT Comparison:}
We measured the average \textit{JCT} to evaluate the performance.
To be more specific, \textit{JCT} is measured from the point of arrival in the frontend scheduler to the point when the prompt's response has been completely formed and stored in the frontend scheduler.
We average the \textit{JCT} of 200 prompts sampled from LMSYS-Chat-1M data.
To provide a fair comparison, we use the same set of sampled prompts but randomly shuffle them for each experiment and repeat this process for three iterations.
Each experiment is conducted with a multiple of the average request rate, which is calculated with the equation listed below (using the average latency listed in Table~\ref{tab:ml-models}):

\begin{figure}[t]
    \centering
    \includegraphics[width=1.0\linewidth]{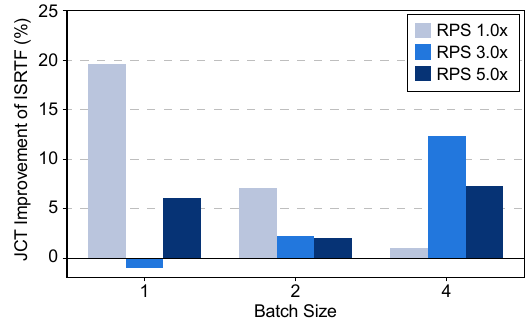}
    \caption{\textit{JCT} improvement of ISRTF over \textsc{FCFS}.}
    \label{fig:improvement_of_ISRTF}
\vspace{0.2in}
\end{figure}
 
\begin{center}
 $AVG. Request Rate$ $=\{\frac{1000}{AVG. Latency}\} \times batch size$
\end{center}

Figure~\ref{fig:aver_jct_per_model} reports the minimum, average, and maximum \textit{JCT} of each experiment.
For almost all cases, \textit{\schedname} has shown enhanced performance by an average of 7.36\% and a maximum of 21.40\%.
This is mostly due to our predictor assigning higher priority to prompts with fewer remaining tokens.
By executing prompts with fewer remaining tokens earlier, \sysname prevents head-of-line blocking as much as possible, reducing queuing delay in the process.
The ideal performance delivered by \textsc{SJF} is listed in Table~\ref{tab:b4_results}.

\paragraph{Deeper Look into Performance Advantages:}
We take a deeper look into a certain case to showcase the main cause of the performance advantage of our result.
The right-hand side figure in Fig.~\ref{fig:aver_jct_per_model} reports the average \textit{JCT} and the average queuing delay of the best case (which is the performance of LlaMA2-13B with a 5.0x RPS.) reported in the left-hand figure, 
The average \textit{JCT} of \textit{\schedname} is 16.45\% lower than that of \textsc{FCFS}.
Additionally, the queuing delay of \textit{\schedname} is 16.75\% smaller than the delay of \textsc{FCFS}.
There is only a 0.30\% difference between the two numbers.
Hence, we can conclude that the major cause of reduced \textit{JCT} is reduced queuing delay.

Additionally, the average scheduling overhead, including batching and the initial BERT predictor, is only 11.04\,ms.
Given the fact that the average latency of LlaMA2-13B (\abb{lam13}) is 8610.2\,ms, the overhead is marginal, accounting for only 0.13\%.
Therefore, it is safe to conclude that the advantage of reduced queuing delay outweighs the performance penalty introduced by additional scheduling overhead.

\begin{figure}[t]
    \centering
    \includegraphics[width=1.0\linewidth]{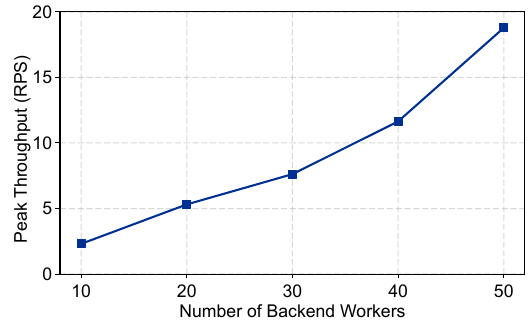}
    \caption{Peak request rate where the average queuing delay of each worker does not exceed \textit{0.5 s} with different number of backend workers.}
    \label{fig:scale_test}
\end{figure}

\subsection{JCT Improvement over Different Batch Sizes}
To provide a comprehensive evaluation, we perform the same experiment introduced in Section~\ref{sec:eval_perf_comparison} with smaller batch sizes, \textit{1} and \textit{2}.
Due to limited space, we omit detailed results but instead provide summarized performance of \schedname for each batch size and RPS multiple compared to \textsc{FCFS}.

Figure~\ref{fig:improvement_of_ISRTF} reports the amount of improvement (in percentage) compared to the performance of \textsc{FCFS} in terms of average \textit{JCT} of \textit{\schedname}.
For example, since the \textit{JCT} enhancement for batch size 1 on RPS 1.0x is 19.58\%, the \textit{JCT} of \textit{\schedname} will be 4.90 seconds less if the \textit{JCT} of \textsc{FCFS} is 25 seconds.
For all (except one) setups, \textit{\schedname} has outperformed \textsc{FCFS}.
We have observed that for experiments with lower batch sizes and high RPS multiples, the chances of \textit{\schedname} reporting higher \textit{JCT} increase.
This phenomenon is due to the larger number of queued prompts mitigating the effect of priority scheduling and making the performance more dependent on the system's throughput itself.

\subsection{Scalability Evaluation}
\label{sec:evaluation_scalability}

To evaluate whether the scheduler and load balancer of \sysname can successfully scale the number of backend workers, we measure the performance of our system by varying the number of workers and incoming request rate.
We measure and report the \textit{peak throughput}, which is the maximum request rate where the average queuing delay of each request does not exceed 0.5 seconds.
Unlike conventional evaluating methods for \textit{peak throughput}, we do not use a direct comparison of latency (i.e., average or P99) when measuring the overhead.
This is because the latency of each prompt is not fixed and changes due to the auto-regressive nature of the decoding process.
Instead, we use the queuing delay as an indicator of scalability since the delay includes the overhead of choosing among multiple workers and waiting time caused by a limited number of backend workers.
Each backend worker handles requests with a maximum batch size of 4 for the LlaMA2-13B model, and the \textit{\schedname} scheduler was used for evaluation.

To assess the scalability of \sysname beyond our limited experimentation environment, we have experimented with \sysname on a larger cluster with NVIDIA H100 GPUs and we deploy one worker per GPU.
Figure~\ref{fig:scale_test} reports the achieved \textit{peak throughput} for each number of backend workers.
We have increased the number of backend workers up to 50 by incrementing 10 workers at a time.
Starting from 2.31 RPS for 10 backend workers, \sysname displayed near-linear scaling performance, achieving 18.77 RPS for 50 workers.
This is due to two reasons:
The first reason is that the load balancer successfully distributed prompts among multiple workers in an efficient manner.
The second reason is that the sub-procedures of scheduling are executed asynchronously, as described in Section~\ref{sec:scheduler}, giving a boost to scalability.

\begin{table}[]
\caption{Avg \textit{JCT}~(s) of each model and scheduling method. Experimented on NVIDIA A100 GPUs with batch size of 4.}
\small
\centering
\begin{tabular}{lllll}
\hline
\textbf{Model}                   & \textbf{RPS}  & \textbf{FCFS}   & \textbf{ISRTF}  & \textbf{SJF}    \\ \hline
\multirow{3}{*}{\textbf{opt13}}  & 1.0x & 77.83  & 73.57  & 20.35  \\
                        & 3.0x & 116.46 & 98.74  & 43.63  \\
                        & 5.0x & 118.13 & 118.11 & 43.63  \\ \hline
\multirow{3}{*}{\textbf{opt6.7}} & 1.0x & 45.08  & 50.52  & 13.21  \\
                        & 3.0x & 83.42  & 72.33  & 24.62  \\
                        & 5.0x & 73.93  & 74.41  & 31.91  \\ \hline
\multirow{3}{*}{\textbf{vic}}    & 1.0x & 93.42  & 73.43  & 32.34  \\
                        & 3.0x & 134.96 & 118.22 & 58.39  \\
                        & 5.0x & 144.23 & 131.38 & 60.98  \\ \hline
\multirow{3}{*}{\textbf{lam13}}  & 1.0x & 240.25 & 212.60 & 70.55  \\
                        & 3.0x & 350.55 & 352.53 & 133.11 \\
                        & 5.0x & 451.59 & 377.29 & 125.59 \\ \hline
\multirow{3}{*}{\textbf{lam7}}   & 1.0x & 91.28  & 130.71 & 37.02  \\
                        & 3.0x & 229.64 & 200.34 & 59.37  \\
                        & 5.0x & 251.66 & 234.08 & 89.64  \\ \hline
\end{tabular}.
\vspace{0.1in}
\label{tab:b4_results}
\vspace{-0.2in}
\end{table}

%% file: 7_related_work.tex

\section{Related Work}
Over the years, numerous machine learning serving systems have been introduced in both industry and academia.
We provide a brief overview of this research and highlight how \sysname differs from each of them.

\paragraph{LLM Response Prediction:}
We first introduce contemporary works related to predicting the response time of LLMs, starting with studies that have also used BERT models.
S3 trained a DistilBERT model to predict the priority of each task~\cite{nips:s3}.
Qiu \textit{et al.} also conducted research on predicting the remaining number of tokens by leveraging the natural language understanding capability of BERT models~\cite{aiops2024qiu}.
While S3 and Qiu \textit{et al.}\ both utilize BERT models, both studies rely on a single prediction and do not provide any backup plans when the prediction is wrong.
In contrast, \sysname predicts priority iteratively and displays stable predictions as each iteration proceeds.

Other studies have also been presented which predict priorities without leveraging a BERT model.
FastServe presented a priority scheduler by deploying a MLFQ and prioritizes tasks with the shortest remaining time in an adaptive manner~\cite{arxiv:fastserve}.
Our work differs from FastServe since we predict the priority rather than adapting and adjusting the priority of a task.
By predicting the priority, we save the overhead caused by the process of finding the priority.
PiA performed instruction-tuning on an LLM by instructing it to predict the response length before generating a response~\cite{nips:PiA}.
Instruction-tuning LLMs may lead to decreased performance and accuracy; in several cases, it has been observed that the length of the actual response is limited to the initial prediction.
This is not an issue for \sysname since we do not use the same model for predicting and generating the response.

\paragraph{LLM Serving Systems:}
We also provide a brief comparison to works related to LLM serving that focus on topics other than response prediction.
Several papers have presented how to optimize the use of the KV cache~\cite{SOSP:vLLM, osdi:infinigen, atc:gao}, while other papers focus on how to load-balance the workload for each worker node in a distributed environment~\cite{osdi:llumnix, asplos:spot_serve, osdi:serverlessllm}.
Additionally, a number of studies have researched how to improve performance by efficiently decoupling the prefill and decode phases~\cite{osdi:sarathi_serve, asplos:exegpt, asplos:specinfer}.
However, the works mentioned above lack one or more of the main contributions of \sysname: (1) scaling to multiple nodes, (2) considering preemption, and (3) analyzing real-world data.

\paragraph{Optimizing Machine Learning Inference:}
Although our work focuses on LLMs, we introduce several categories of research related to optimizing the inference process of general machine learning models.
Such previous works have inspired our work and can be categorized into: scaling a serving system in a cloud environment~\cite{atc:infaas, atc:mark}; reducing the problem space of serving heterogeneous models and goals~\cite{osdi:gavel, sosp:nexus}; researching various parallelisms~\cite{osdi:alpaserve, nips:tensorflow_serving, nsdi:clipper}; and leveraging special features provided by the GPU~\cite{atc:choi, socc:gslice, eurosys:jeong}.

%% file: 8_conclusion.tex

\section{Conclusions}
We present \sysname, an \textbf{E}fficient \textbf{L}LM \textbf{I}terative \textbf{S}cheduling system, which leverages the natural language understanding capabilities and relatively low computation cost of BERT model for priority scheduling.
Based on the predictions of our model, we prioritize tasks that have fewer predicted remaining tokens to minimize head-of-line blocking and enhance performance.
Not only have we tailored a BERT model for prediction, but we also provide an analysis of real-world data obtained by observing requests in \svcname.
The prototype system is deployed as Kubernetes components, a widely used container orchestration platform in cloud industry, to facilitate future research and deployment of our system.
Experimental results demonstrate that the iterative scheduling method deployed in \sysname reduces average JCT by up to 19.6\%.

%% file: 9_ack.tex
\section{Acknowledgments}
We extend our gratitude to the \compname AI research team for their advice on model training and \compname \svcname team for providing actual LLM service data.

%% file: N_Appendix.tex
\begin{appendices}
\section{Preemption Profiling}
\label{sec:append_data}

We present our profiling data obtained from investigating the probability of preemption due to limited memory space.
This section includes the profiling setup and the minimum batch size that causes a preemption.
All profiling was conducted on our prototype \sysname backed by NVIDIA A100 GPUs.
We profiled each model listed in Table~\ref{tab:ml-models} with 10K prompts that were sampled from the LMSYS-Chat-1M dataset~\cite{zheng2023lmsyschat1m}.
To simulate an environment where there are sufficient prompts to form large batches, we saturated the job pool in the frontend scheduler with a high request rate (10K requests per second).
We gradually increased the batch size by 10, up to a maximum of 250, until we observed a preemption.
If a preemption was not observed, we reduced the memory utilization of vLLM (from the default 90\%) and repeated the process.
All profiling results are reported in Table~\ref{tab:prempt_profile_result}.

\begin{table}[ht!]
\caption{Profiling results for preemption. \textit{Batch size} represents the minimum batch size at which a preemption has occurred.}
\begin{tabular}{lcc}
\hline
\textbf{Model Name} & \multicolumn{1}{l}{\textbf{Batch Size}} & \multicolumn{1}{l}{\textbf{vLLM Memory Limit}} \\ \hline
\textbf{LlaMA2-13B} & 120                                      & 90\%                                                 \\
\textbf{LlaMA2-7B}  & 40                                       & 30\%                                                    \\
\textbf{OPT-7B}     & 30                                       & 40\%                                                  \\
\textbf{OPT-13B}    & 60                                       & 40\%                                                    \\
\textbf{Vicuna-13B} & 90                                       & 40\%                                                    \\ \hline
\end{tabular}

\label{tab:prempt_profile_result}
\end{table}

\section{Models Used for Training}
We provide the list of models used for training the predictor presented in Section~\ref{sec:predictor}.
Table~\ref{tab:13model_list} reports the size of each model and the organization that produced it.

\begin{table}[ht]
\caption{List of models for training.}
\begin{tabular}{ccl}
\hline
\multicolumn{1}{l}{\textbf{Model Name}} & \multicolumn{1}{l}{\textbf{Size (B)}} & \textbf{Producer}\\ \hline
\textbf{LlaMA-7B}                       & 7                                            &Meta\\
\textbf{LlaMA-13B}                      & 13                                           &Meta\\
\textbf{LlaMA2-7B}                      & 7                                            &Huggyllama\\
\textbf{LlaMA2-13B}                     & 13                                           &Huggyllama\\
\textbf{Vicuna-7B}                      & 7                                            &LMSYS\\
\textbf{Vicuna-13B}                     & 13                                           &LMSYS\\
\textbf{OPT-1B}                         & 1.3                                          &Facebook\\
\textbf{OPT-3B}                         & 2.7                                          &Facebook\\
\textbf{OPT-7B}                         & 6.7                                          &Facebook\\
\textbf{OPT-13B}                        & 13                                           &Facebook\\
\textbf{GPT-NeoX}                       & 20                                           &EleutherAI\\
\textbf{Gemma}                          & 7                                            &Google\\
\textbf{SOLAR}                          & 11                                           &Upstage\\
\hline
\end{tabular}
\label{tab:13model_list}
\end{table}

\end{appendices}